\documentclass[conference]{IEEEtran}

\usepackage{ucs}
\usepackage[utf8x]{inputenc}
\usepackage[T2A,T1]{autofe}

\usepackage{makeidx}
\usepackage{graphicx}
\usepackage{stringenc}
\usepackage[filenameencoding=utf8x]{grffile}
\usepackage{hyperref}

\newcommand{\noitem}[1][]{{\it #1}}

\title{$\pi$-Control: A Personal Cloud Control Centre}
\author{\IEEEauthorblockN{Josef Spillner}
\IEEEauthorblockA{Faculty of Computer Science\\
Technische Universität Dresden\\
01062 Dresden, Germany\\
Email: josef.spillner@tu-dresden.de}
\and
\IEEEauthorblockN{Alexander Schill}
\IEEEauthorblockA{Faculty of Computer Science\\
Technische Universität Dresden\\
01062 Dresden, Germany\\
Email: alexander.schill@tu-dresden.de}}

\begin{document}

\maketitle

\begin{abstract}
Consumption of online services and cloud computing offerings is on the rise, largely due to compelling advantages over
traditional local applications. From a user perspective, these include zero-maintenance of software, the always-on nature
of such services, mashups of different applications and the networking effect with other users.
Associated disadvantages are known, but effective means and tools to limit their effect are not yet
well-established and not yet generally available to service users.
We propose (1) a user-centric model of cloud elements beyond the conventional <SPI>aaS layers,
including activities across trust zones,
and (2) a personal control console for all individual and collaborative user activities in the cloud.
\end{abstract}



\section{Introduction}

The Internet, originally thought to be a set of peer-to-peer connections between its users, has turned into a set of unequal
participants. On the lower level of wires and wireless connections, the concentration of traffic in backbones
and routers has valid technical reasons. On the higher level of Internet applications, any significant concentration is a symptom of portals,
marketplaces, walled gardens, and the general asymmetric differentiation between producers and consumers of information. This decreases the collaborative
potential of sharing applications, services, data and resources.
The increasing dependency of users on service and cloud computing providers and their reduced leeway is often met with skepticism, although the number
and impact of counter-measures remains low. Users still lack overview and control mechanisms for their digital trust domain,
typically consisting of devices and resources under their control.
This issue is likely to become worse as the mandatory use of government-provided online services for citizens is on the rise.
Without appropriate information and control facilities, the users' informational self-determination will be severely decreased.
Therefore, we propose $\pi$-Control, an abstract personal control centre for all user activities in the cloud. Its architecture is based
on a model of typical cloud elements and workflows. The power of the control centre approach encompasses summaries of current and
historic activities as well as context-aware service provisioning, migration and replication tools.

The remainder of this document introduces the model of cloud elements and activities, the design criteria of $\pi$-Control
derived from identified problems which can be represented in the model, and a proposal for a software architecture
to realise $\pi$-Control as installable software or dedicated appliance.

\section{User-Centric Cloud Elements}

\begin{figure}[h]
\center
\includegraphics[width=0.44\columnwidth]{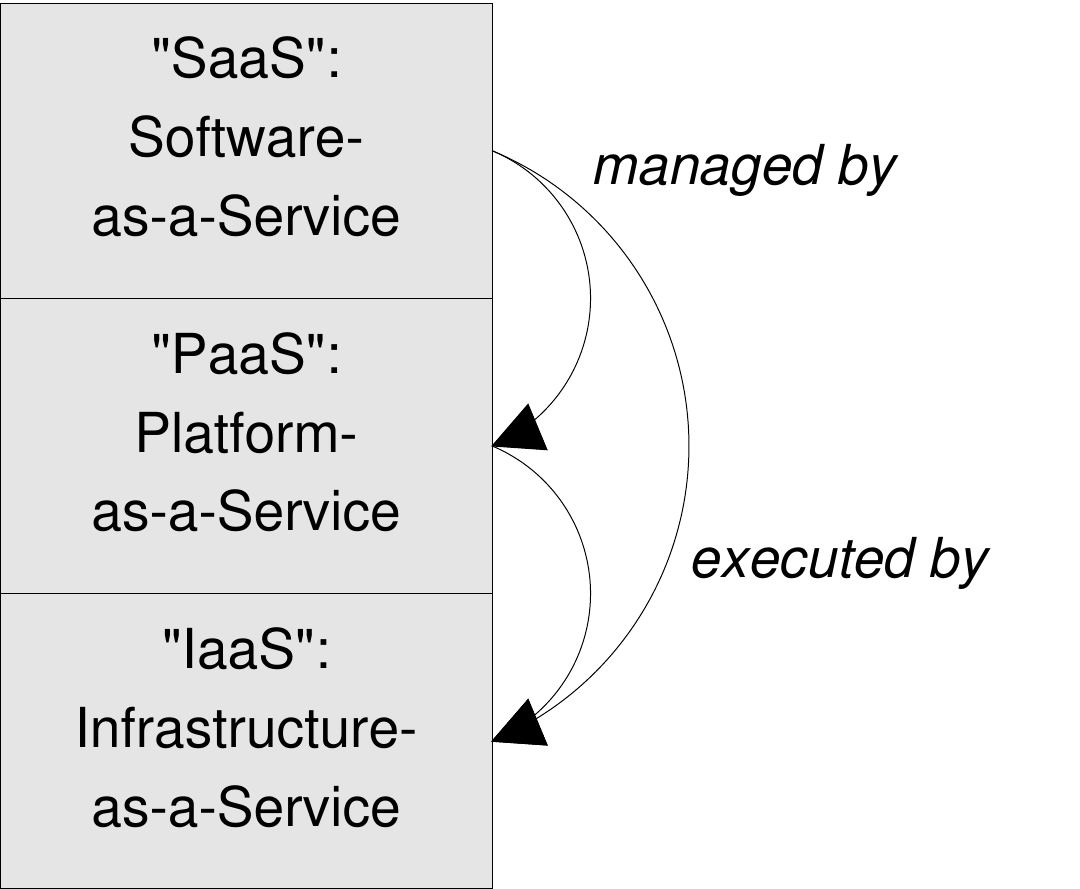}
\caption{Conventional administrative interpretation of cloud layers}
\label{fig:cloudlayers}
\end{figure}

The digital world consists of a variety of objects accessible as services and interpretable in various ways, e.g. usage,
modification operations and execution. A common view of the elements of a cloud computing architecture is given in Fig.
\ref{fig:cloudlayers}. Software services are managed by platform services, with both parts being executed on infrastructure
also offered as a service. The view has been defined as the <SPI>aaS cloud model \cite{nistcloud}.

This administrator-centric layering of cloud elements is less suitable for the consideration of user-centric access to the cloud.
The central reason for the discrepancy is that users are typically not interested in explicit invocation of the platform layer which
rather serves as transparent
middleware, offering functions such as login, deployment, search and monitoring reports. The platform is also inherently more static than
tradeable software and infrastructure services which are contracted by users explicitly based on their non-functional properties such as
price or legal status.
Therefore, we propose
a different interpretation, taking into account a wider service definition from recent research objectives such as
data clouds \cite{relationalcloud} and the reduction of risks associated with cloud computing \cite{abovethecloud}.

For simplicity, we start with a definition of $O := \{S,D,R,...\}$ in which the globally accessible service-oriented offering
of objects $O$ consists of
(executable) software $S$, (non-executable) data $D$ and resources $R$. The set of service domains of high interest to the user is kept
open on purpose to allow for future additions.
There is a runtime dependency of software on computing resources and
almost always on data, as well as a permanent dependency of both software and data on resources for storage.
The cloud computing terms SaaS, DaaS and RaaS add service interfaces to software, data stores and resources, respectively.
Hence, $O$ represents digital objects in a service-oriented manner.

\begin{figure}[h]
\center
\includegraphics[width=\columnwidth]{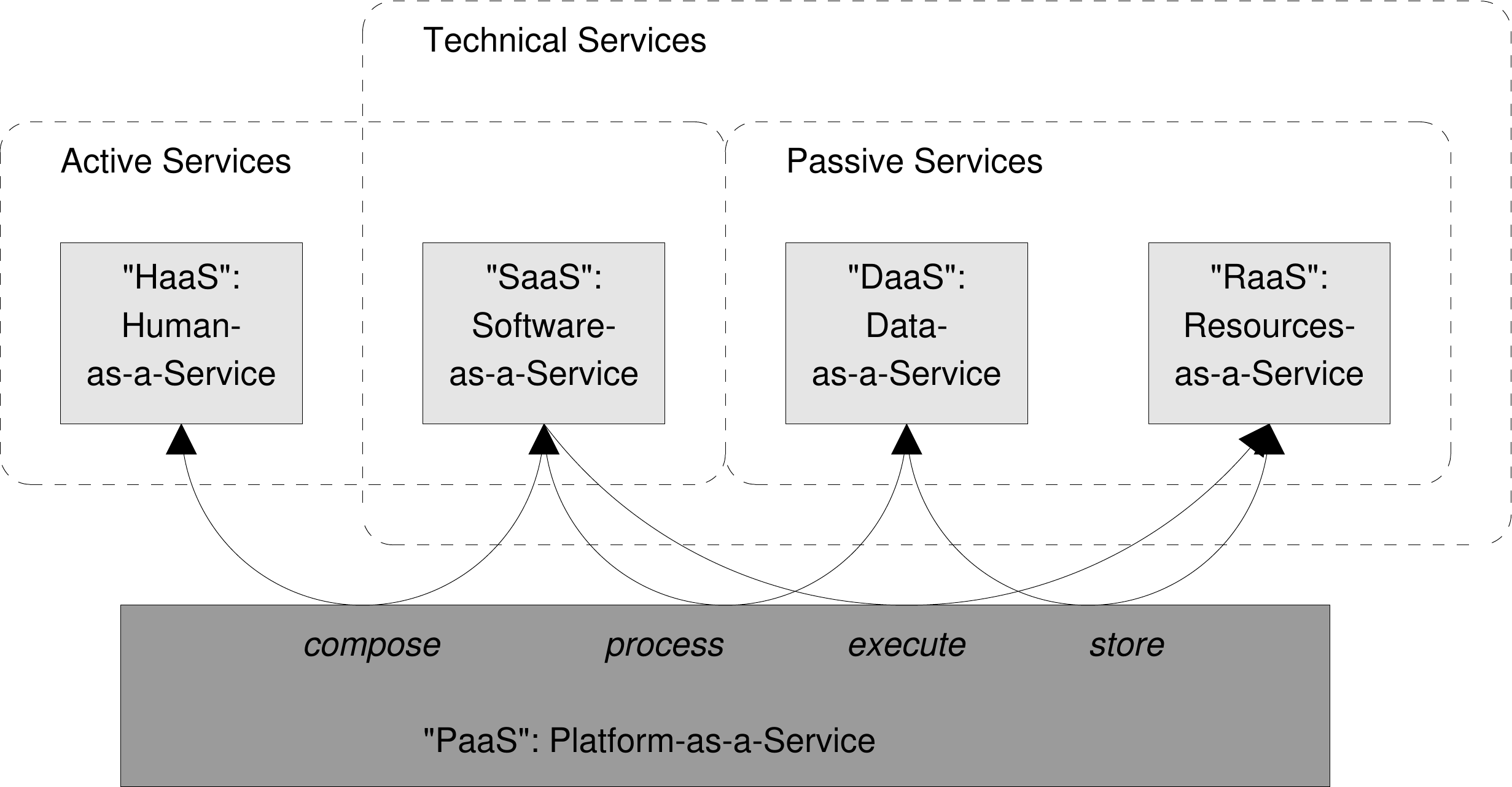}
\caption{Service classes in the cloud from a user perspective: Boundaries, characteristics and dependencies}
\label{fig:servicedefinition}
\end{figure}

This user-centric service definition for cloud computing scenarios is shown in Fig. \ref{fig:servicedefinition}.
We believe that it is not contradicting previously consolidated comprehensive cloud model definitions \cite{cloudbreak}.

The exclusion of the PaaS-level service platform, possibly still consisting of loosely-coupled and uniform platform
services, from services of interest to the user puts the platform into a more hidden, yet central and
ubiquitous position. It allows for a transparent distribution of the platform, e.g. by connecting several devices,
and a transparent aggregation of various concrete platforms into a logical one.
Furthermore, it assumes from the beginning that at least one instance of the platform services be under the user's control.

The proposed <SDR>aaS cloud service definition also conveys the von Neumann model for computers better than the conventional
<SPI>aaS model. The separate treatment of software
and data services, both in combination with supportive resource services and complementary human services,
will lead to flexible views on distributed workflows and scale-out scenarios for software applications.

\section{User-Centric Cloud Activities}

User activities in distributed cloud computing scenarios usually evolve around personal usage (e.g. device synchronisation, backup) and collaborative
service use on community-level restricted or public open directories and marketplaces (e.g. meeting time planner). This introduces the implicit
and subjective notion of trust zones with most users leaning towards entrusting all data to resources on devices under their
control, most data to resources under control of recognised friends or colleagues, some data to resources of identified
companies and operators, and only uncritical data to completely unknown resources.

Directories and marketplaces for digital objects can usually be categorised along the same definition of <SDR>aaS by having $M(R)$, $M(S)$ and $M(D)$
as logically separate (albeit possibly physically combined) brokers for each kind of service.
Digital objects can be replicated into personal domains $M_{T0}(O)$ in the user's trust zone and under the user's control.
Objects can also be replicated from there to other domains including community directories $M_{T1}(O)$ or again public
free, commercial and governmental marketplaces $M_{T>=2}(O)$. The trust level definition is a subjective metric whose only purpose is to define
a total order of preference for migration and replication strategies.
Resources cannot be replicated. They can either be migrated physically, or access to them can be migrated
(wholly or partially, as in sharing) as virtual digital objects.

\section{Potential Problem Identification}

Under the assumptions of the model of cloud elements and activities presented in the previous two sections,
three major potential problems can be identified which, in their essence, also apply to other cloud computing models.
Design criteria for a personal cloud control architecture should be set in a way that they prevent them
from turning into actual problems and threats to the user.

\noitem[Users lack information.] Without appropriate real-time and historic information about relations to
service providers and access to data, users will not be able to make the right decisions.
For any amount of $O$, users should be able to keep an overview about elements and activities in
$M(O)$ irrespective of the trust domain.
The completeness of the overview should be total for $T=0$ and may become less for larger $T$.

\noitem[Users lack control.] Users should also be informed about the context-dependent characteristics of $M_{T>0}(O)$
so they can start replicating, migrating, providing and sharing their objects accordingly.
Without powerful control tools, the user might be locked into certain cloud environments while high costs are associated
with any attempt at reversing this situation \cite{csal}.

\noitem[Users lack autarky.] Just like users' private data should be under their control, there should be mechanisms to replicate
public data sets and appropriately licenced public services into the local domain. Hence, a transformation of $T>0$ to $T=0$ is
required.
This especially applies to data on the platform level, such as the contents of a service registry, so that a
non-trivial amount of cloud control tasks can be initiated even when being temporarily offline.
Solving this problem also means giving users the tools to provide services by themselves, if needed.



\section{Existing Approaches}


While the problems extracted in the previous section have already been known to some extent and for some time,
existing approaches to solve them do not focus on their combination.

Private Virtual Infrastructures, representing a new cloud management model \cite{trustedutilitycloud}, shift security and privacy
risks midway back to the provider and thereby reduce efforts required by the user. However, they assume changes to today's IaaS
and do not consider the <PS>aaS layers and hence the interactive involvement of the user, turning them more into a
complementary base technology. Nevertheless, their embodied secure migration processes could serve as a realisation
of the corresponding cloud activities.

Personal software and data distribution systems, which are increasingly integrated into operating system desktops,
provide sophisticated repository and peer-to-peer sharing, versioning, and dependency control. An example is the
advanced Debian package solver \cite{distributioncomplexity}. However, these systems
currently lack integration with resource control and user management systems.
Another approach is to control the user's activity mobility in certain collaborative contexts \cite{activitycontrol}.
It is targeting interactive sessions in distributed operating systems rather than heterogeneous cloud computing environments.

The definition of trust in cloud environments is a fairly new research topic \cite{cloudtrust}. Therefore, our work
omits further formalisation of this aspect and relies on a hierarchical scalar trust metric.

\section{Proposed Software Architecture}

Built upon the user-centric model of cloud elements and activities, and influenced by the goal to overcome weaknesses
of related approaches to solve the identified potential problems, we suggest an abstract cloud control architecture.
It is supposed to be used as a blueprint for an interactive realisation
for exerting control of the personal service provisioning and consumption activities in public and community cloud computing
environments.

\begin{figure}[h]
\center
\includegraphics[width=0.62\columnwidth]{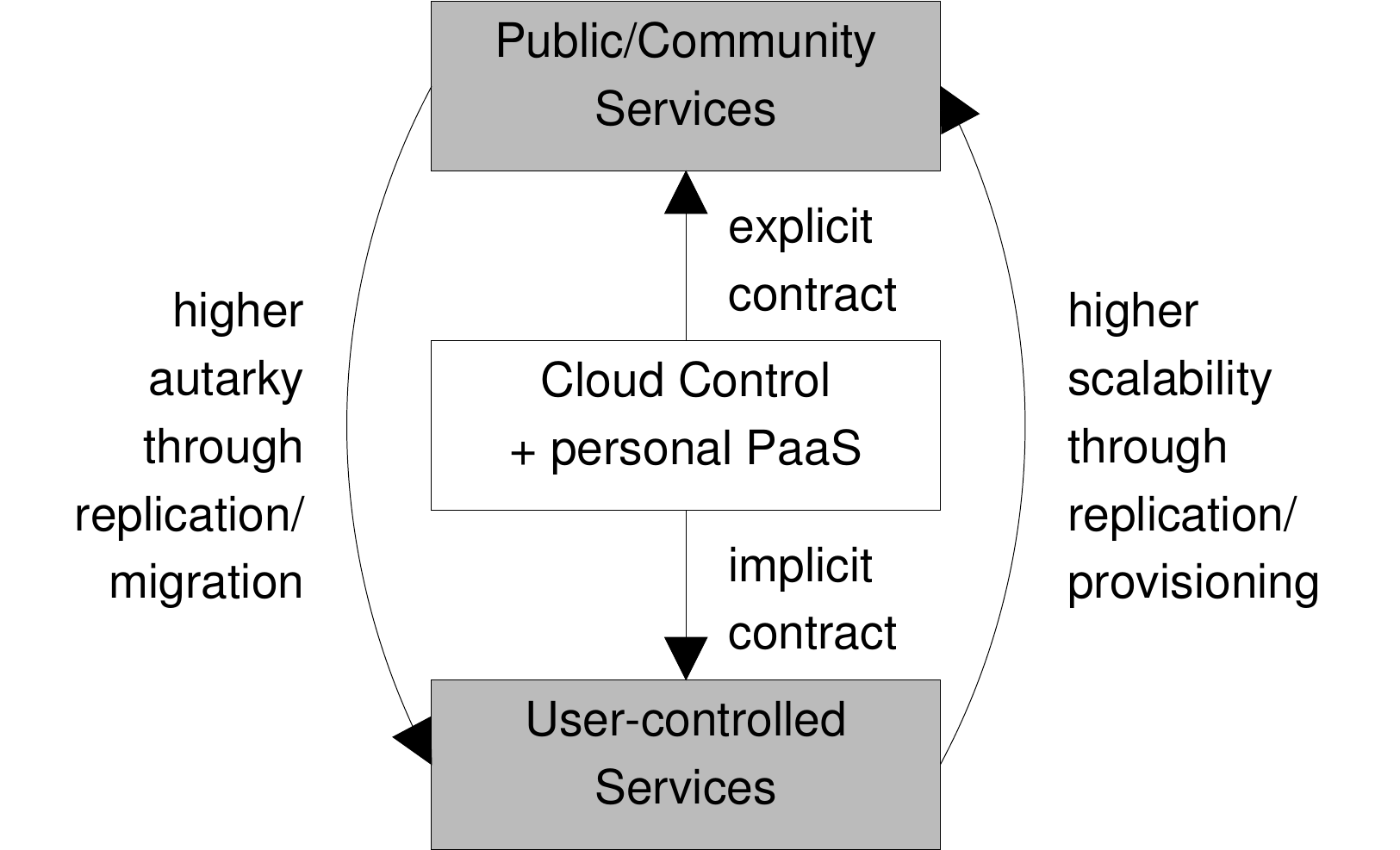}
\caption{$\pi$-Control environment}
\label{fig:picontrolenv}
\end{figure}

The proposed personal cloud control functionality, named $\pi$-Control,
imports public lists of objects available on marketplaces $M_{T>0}(O)$ by their
respective category of $S$, $D$ and $R$. Matching private directories of objects available to the user $M_{T0}(O)$
are managed similarly. Users can search in all directories and extend them by advertising their own objects, via conventional
link-only registration or export of the object itself.
The environment surrounding $\pi$-Control is shown in figure \ref{fig:picontrolenv}.

\begin{figure}[h]
\center
\includegraphics[width=\columnwidth]{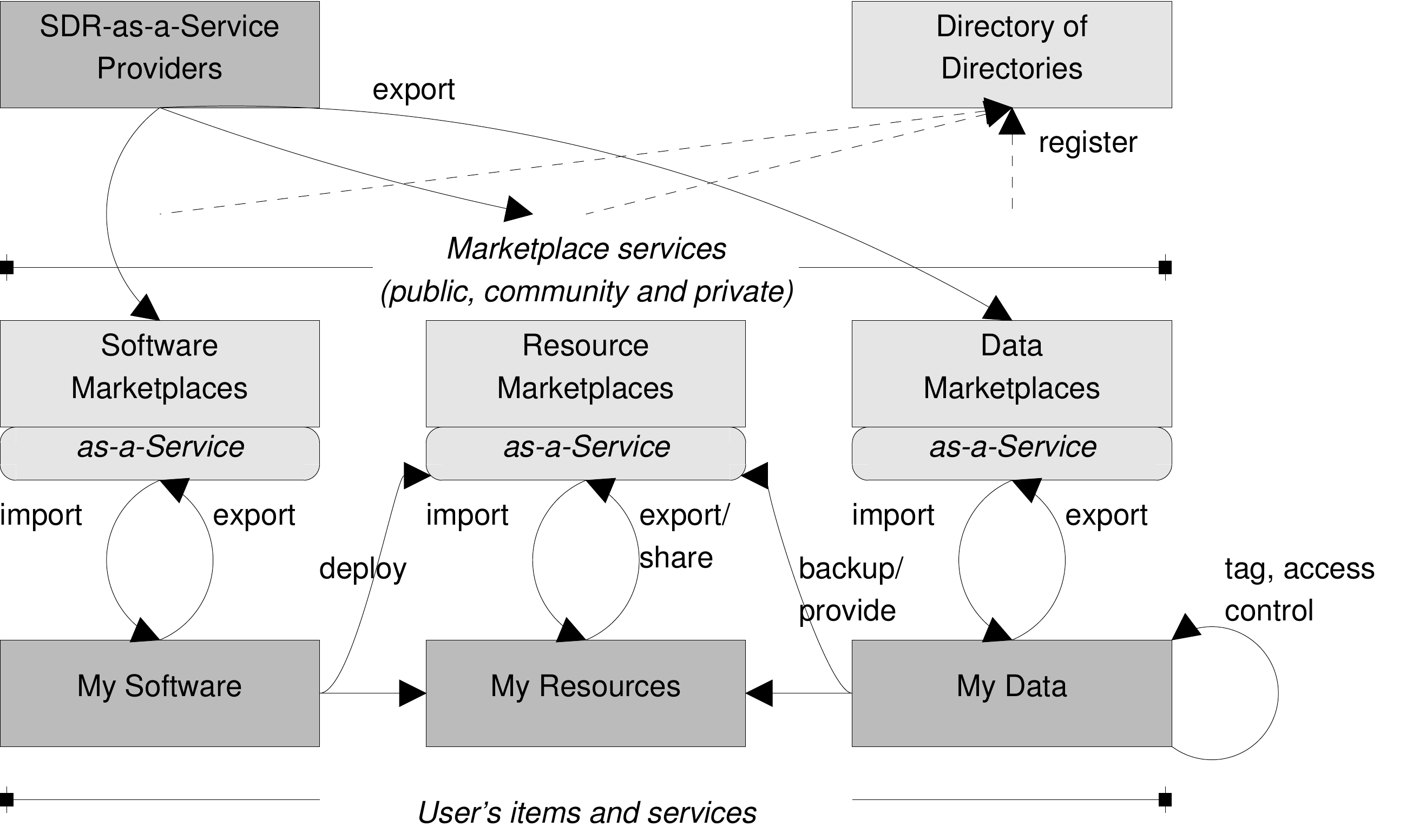}
\caption{$\pi$-Control architecture}
\label{fig:picontrol}
\end{figure}

Users shall have the possibility of not just migrating and replicating the objects between
trust zones, but also using them in various ways. This includes the deployment of software and storage of data onto resources.
Furthermore, users should be able to manage their data, including tagging and access control to influence data
placement and external use strategies.
Given that $\pi$-Control interacts with a PaaS in the user's trust domain, service contracting through SLAs and
non-guaranteed property descriptions will direct and constrain the usage options. In addition, a local PaaS can be
used to empower the user to provide services without relying on untrusted infrastructure.
The resulting abstract architecture is shown in Fig. \ref{fig:picontrol}.

Service directories are based upon public marketplace interfaces, community registries, local network auto-discovery services, and
strictly local information depending on the device on which $\pi$-Control runs.

For each new directory and service provider, a trust level can be assigned, with trust inheritance from directories
to providers and from providers to services unless overridden by the user. Directories of directories can similarly
be added to find out about new markets which, in turn, inherit the directories' trust levels.
In parallel to the user-defined trust, author-defined licencing rules govern the general use of data and software
services. For example, replicas (or partial caches) of public open data sets from governments or researchers
will generally be made available to anybody with login permission to the $\pi$-Control-governed devices.

Two kinds of context will influence the control centre behaviour: contracts with service providers, and their availability.

Several activities are contract-context-dependent. For example, backup of data will need at least one storage service contract.
With at least two such contracts, backups can be split and redundantly dispersed with secret-sharing algorithms, lowering the
dependency on individual providers.
Likewise, outsourcing services through remote deployment requires a contract with a compute service.

Other activities are availability-context-dependent. For example, a nearby storage server might not always be switched on
or not available when roaming. Similarly, a contract-bound data provider might not be reachable. Data replication
and synchronisation techniques working on platform-level data like registry information and on DaaS-level data
minimise the impact of limited availability.

In order to achieve higher autarky and more precise information, we propose to rely on distributed version control systems.
These systems combine the advantages of peer-to-peer systems, including even offline operation, with reliable history
information and roll-back capabilities.

All services need to be sufficiently described. Notable attributes include identification, function, provider information,
pricing, further non-functional properties and technical requirements. New developments in service descriptions such
as the Unified Service Description Language (USDL) promise to solve this task. However, USDL has only been evaluated
for SaaS and Human-as-a-Service (HaaS) so far, not yet for RaaS and DaaS. The level of automation which can be
achieved for service placement and data replication strategies is closely tied to the expressivity, accuracy
and general quality of the property specifications.

User privileges, access permissions, roles and identities have historically been implemented within the applications.
As the trend towards SaaS continues, these concerns are increasingly confined to an appropriate service structure.
However, many cloud providers run their own user management infrastructure. As a requirement especially for collaborative
scenarios, such as smart office applications delivered from the cloud for dynamically composed groups of users,
$\pi$-Control shall treat user privileges as a dedicated data set which is subject to the same replication mechanisms
as regular sets. This way, configurations of access control to multi-tenant services can easily be propagated to various cloud providers.
The access control rules are applied in conjunction with the object licencing metadata. An existing use case covered
by this combination is the collaborative work on public open data which is tagged as such \cite{collaborativecaches}.

To summarise the architectural concept: Based on rich and high-quality declarative service descriptions and integration with
local PaaS-level service provisioning infrastructures, the control centre offers synchronised,
context-dependent and service-kind-dependent overview and control functionality for individual and collaborative
cloud computing usage scenarios.

\section{User Interface Considerations}

Considering that the target groups of $\pi$-Control are consumers and producers in the cloud, as opposed to experienced
operators, the user interface should be clean, intuitive and free of unexpected surprises. We believe that there is room
for new interaction patterns, such as drag-and-drop for service and data migration between clouds, hiding the migration
details and underlying protocols.

\begin{figure}[h]
\center
\includegraphics[width=\columnwidth]{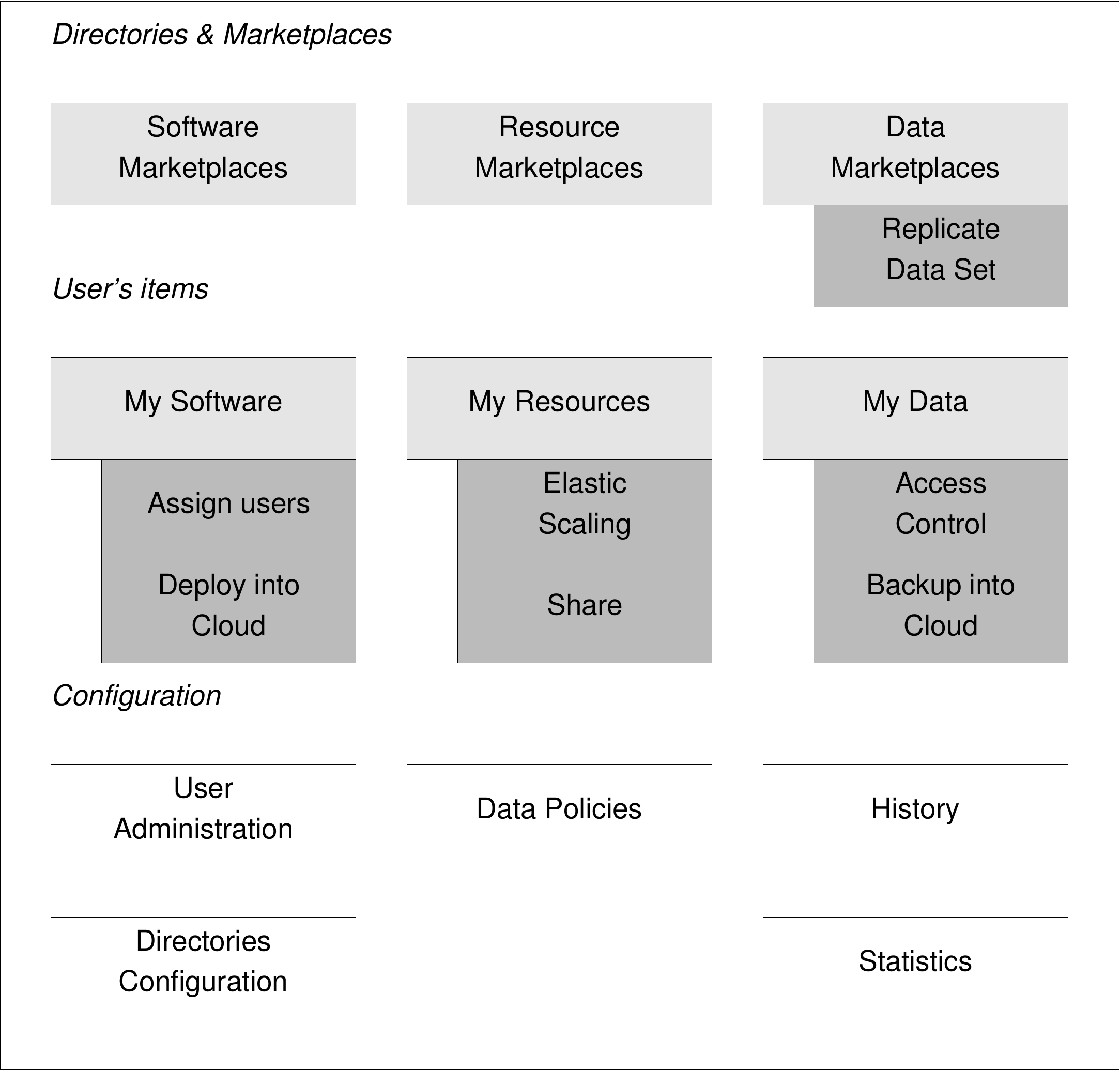}
\caption{$\pi$-Control user interface mockup}
\label{fig:picontrolui}
\end{figure}

A sketch for one possible variant of the user interface is given in Fig. \ref{fig:picontrolui}.
It clearly differentiates between offers from directories in various trust domains and objects under the control
of the user, including their instances and context-dependent actions.

\section{Summary and Outlook}

We have discussed the need for personal cloud control centres and introduced a suitable architecture based on
a custom model of cloud elements and activities.
Within the next months, we will work on an implementation within the context of the
FlexCloud\footnote{FlexCloud: \url{http://flexcloud.eu/}. This work has received from the European Social Fund and the Free State of Saxony, Germany, under project number 080949277.}
project. Special attention will be paid to the consideration of non-functional properties
for service placement and data replication strategies. Moreover, we plan to build an appliance consisting of
hardware resources, a preconfigured service platform installation and $\pi$-Control running on top of it.

\bibliographystyle{IEEEtran}
\bibliography{picontrol}

\begin{thebibliography}{10}
\providecommand{\url}[1]{#1}
\csname url@samestyle\endcsname
\providecommand{\newblock}{\relax}
\providecommand{\bibinfo}[2]{#2}
\providecommand{\BIBentrySTDinterwordspacing}{\spaceskip=0pt\relax}
\providecommand{\BIBentryALTinterwordstretchfactor}{4}
\providecommand{\BIBentryALTinterwordspacing}{\spaceskip=\fontdimen2\font plus
\BIBentryALTinterwordstretchfactor\fontdimen3\font minus
  \fontdimen4\font\relax}
\providecommand{\BIBforeignlanguage}[2]{{%
\expandafter\ifx\csname l@#1\endcsname\relax
\typeout{** WARNING: IEEEtran.bst: No hyphenation pattern has been}%
\typeout{** loaded for the language `#1'. Using the pattern for}%
\typeout{** the default language instead.}%
\else
\language=\csname l@#1\endcsname
\fi
#2}}
\providecommand{\BIBdecl}{\relax}
\BIBdecl

\bibitem{nistcloud}
P.~Mell and T.~Grance, ``{The NIST Definition of Cloud Computing version 15},''
  2009.

\bibitem{relationalcloud}
C.~Curino, E.~Jones, Y.~Zhang, E.~Wu, and S.~Madden, ``{Relational Cloud: The
  Case for a Database Service},'' MIT Computer Science and Artificial
  Intelligence Laboratory, Tech. Rep. MIT-CSAIL-TR-2010-014, March 2010.

\bibitem{abovethecloud}
M.~Armbrust, A.~Fox, R.~Griffith, A.~D. Joseph, R.~H. Katz, A.~Konwinski,
  G.~Lee, D.~A. Patterson, A.~Rabkin, and M.~Zaharia, ``{Above the Clouds: A
  Berkeley View of Cloud Computing},'' UC Berkeley Reliable Adaptive
  Distributed Systems Laboratory, Tech. Rep., February 2009.

\bibitem{cloudbreak}
L.~M. Vaquero, L.~Rodero-Merino, J.~Caceres, and M.~Lindner, ``{A Break in the
  Clouds: Towards a Cloud Definition},'' ACM SIGCOMM Computer Communication
  Review (Editorial Article), pp. 50--55, 2009.

\bibitem{csal}
Z.~Hill and M.~Humphrey, ``{CSAL: A Cloud Storage Abstraction Layer to Enable
  Portable Cloud Applications},'' in \emph{Proc. of 2nd IEEE International
  Conference on Cloud Computing Technology and Science}, December 2010,
  {Indianapolis, Indiana, USA}.

\bibitem{trustedutilitycloud}
F.~J. Krautheim, ``{Building Trust Into Utility Cloud Computing},'' Ph.D.
  dissertation, University of Maryland, Faculty of the Graduate School, April
  2010.

\bibitem{distributioncomplexity}
F.~Mancinelli, J.~Boender, R.~D. Cosmo, J.~Vouillon, B.~Durak, and X.~Leroy,
  ``{Managing the Complexity of Large Free and Open Source Package-Based
  Software Distributions},'' in \emph{Proceedings of the 21st IEEE/ACM
  International Conference on Automated Software Ingineering (ASE)}.\hskip 1em
  plus 0.5em minus 0.4em\relax IEEE Computer Society Press, 2006, pp. 199--208.

\bibitem{activitycontrol}
K.~Leal, F.~Ballesteros, E.~Soriano, and G.~Guardiola, ``{UbiTerm: a hand-held
  control-center for user’s activity mobility},'' in \emph{Proceedings of the
  IEEE International Conference on Pervasive Services}.\hskip 1em plus 0.5em
  minus 0.4em\relax IEEE Computer Society, July 2005, pp. 127--136, {Santorini
  Island, Greece}.

\bibitem{cloudtrust}
S.~S. Thorpe, ``{Modeling a Trust Cloud Context},'' Workshop for Ph.D. Students
  in Information and Knowledge Management (PIKM), October 2010, {Toronto,
  Canada}.

\bibitem{collaborativecaches}
S.~S. Vazhkudai, X.~Ma, V.~W. Freeh, J.~W. Strickland, N.~Tammineedi, T.~Simon,
  and S.~L. Scott, ``{Constructing Collaborative Desktop Storage Caches for
  Large Scientific Datasets},'' \emph{ACM Transaction on Storage}, 2005.

\end{thebibliography}

\end{document}